\documentclass{article}
\usepackage[T1]{fontenc} 
\usepackage[utf8]{inputenc} 
\usepackage{ismir,amsmath,cite,url}
\usepackage{graphicx}
\usepackage{color}

\newenvironment{tight_enumerate}{
\begin{enumerate}
  \setlength{\itemsep}{0pt}
  \setlength{\parskip}{0pt}
}{\end{enumerate}}

\usepackage{amsfonts}

\usepackage{lineno}

\usepackage{xspace}
\usepackage{tcolorbox}
\tcbuselibrary{listings}

\newtcblisting{inlinecode}{
  listing only,
  nobeforeafter,
  after={\xspace},
  hbox,
  tcbox raise base,
  fontupper=\ttfamily,
  colback=lightgray,
  colframe=lightgray,
  size=fbox
  }

\title{VampNet: Music Generation via \\ Masked Acoustic Token Modeling}

\multauthor
{Hugo Flores García$^{1,2}$ \hspace{1cm} Prem Seetharaman$^1$ \hspace{1cm} Rithesh Kumar$^1$ \hspace{1cm} Bryan Pardo$^2$} { \bfseries{}\\
  $^1$ Descript Inc.\\
  $^2$ Northwestern University\\
{\tt\small hugofg@u.northwestern.edu}
}

\def\authorname{H. Flores García, P. Seetharaman, R. Kumar, and B. Pardo}

\makeatletter
\newcommand\footnoteref[1]{\protected@xdef\@thefnmark{\ref{#1}}\@footnotemark}
\makeatother

\begin{document}

\maketitle
\begin{abstract}
We introduce VampNet, a masked acoustic token modeling approach to music synthesis, compression, inpainting, and variation. 
We use a variable masking schedule during training which allows us to sample coherent music from the model by applying a variety of masking approaches (called prompts) during inference. VampNet is non-autoregressive, leveraging a bidirectional transformer architecture that attends to all tokens in a forward pass. With just 36 sampling passes, VampNet can generate coherent high-fidelity musical waveforms. We show that by prompting VampNet in various ways, we can apply it to tasks like music compression, inpainting, outpainting, continuation, and looping with variation (vamping). Appropriately prompted, VampNet is capable of maintaining style, genre, instrumentation, and other high-level aspects of the music. This flexible prompting capability makes VampNet a powerful music co-creation tool. Code\footnoteref{note1} and audio samples\footnoteref{note2} are available online.
    
\end{abstract}
\section{Introduction}\label{sec:introduction}
\begin{figure}
 \centerline{
 \includegraphics[width=1.0\columnwidth]{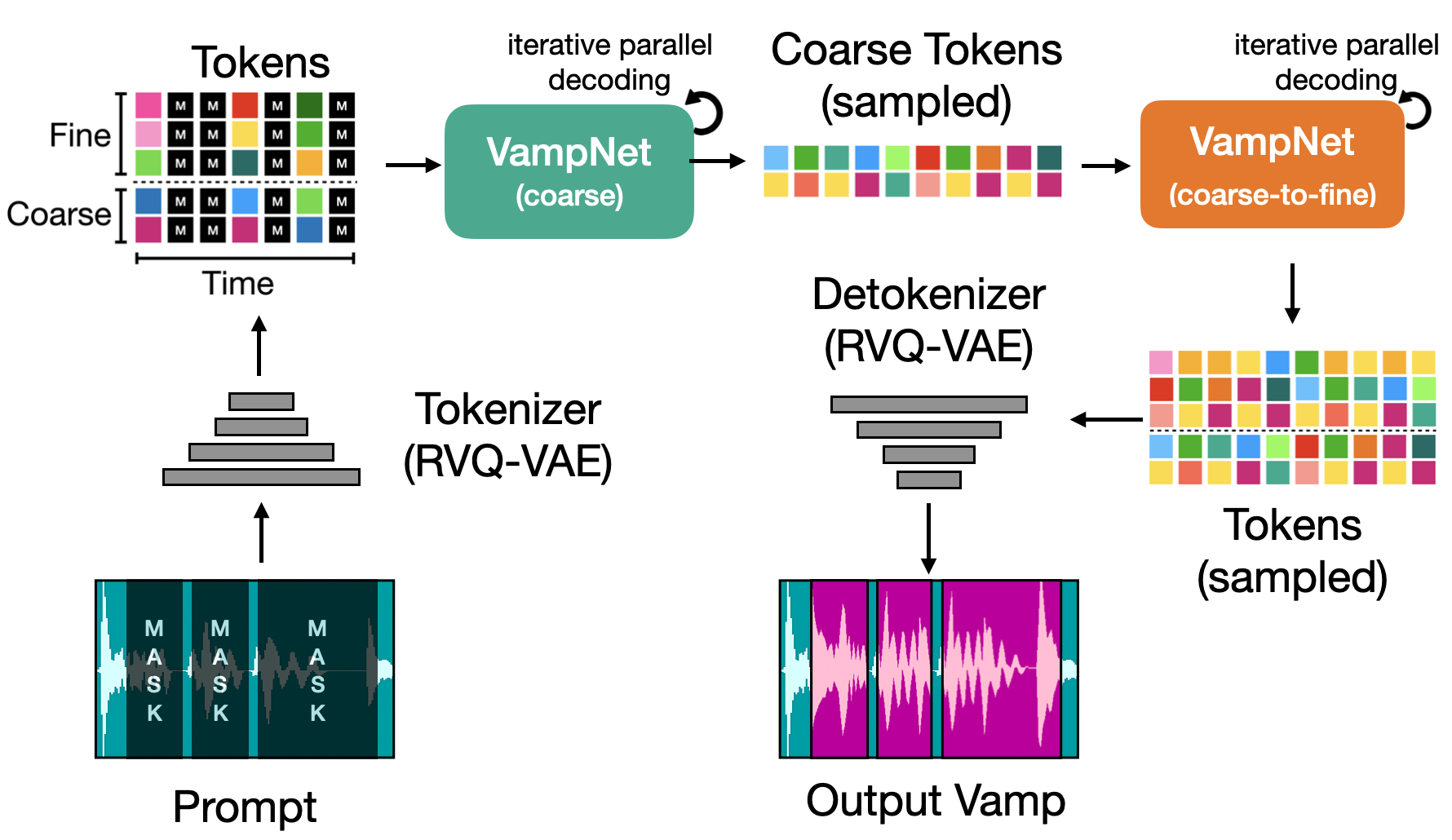}}
 \caption{VampNet overview.  We first convert audio into a sequence of discrete tokens using an audio tokenizer. Tokens are masked, and then passed to a masked generative model, which predicts values for masked tokens via an efficient iterative parallel decoding sampling procedure at two levels. We then decode the result back to audio.}
 \label{fig:inference}
\end{figure}

In recent years, advances in discrete acoustic token modeling have resulted in significant leaps in autoregressive generation of speech \cite{wang2023neural, borsos2022audiolm} and music \cite{agostinelli2023musiclm}. Meanwhile, approaches that use non-autoregressive parallel iterative decoding have been developed for efficient image synthesis \cite{chang2022maskgit, rampas2022paella}. Parallel iterative decoding promises to allow faster inference than autoregressive methods and is more suited to tasks like infill, which require conditioning on both past and future sequence elements. 

In this work, we combine parallel iterative decoding with acoustic token modeling, and apply them to music audio synthesis.  To the best of our knowledge, ours is the first \footnote{While our work was under peer review, Google released SoundStorm \cite{borsos2023soundstorm}, which leverages a similar parallel iterative decoding approach to ours.} extension of parallel iterative decoding to neural audio music generation.  Our model, called VampNet, can be flexibly applied to a variety of applications via token-based prompting. We show that we can guide VampNet's generation with selectively masked music token sequences, asking it to fill in the blanks. The outputs of this procedure can range from a high-quality audio compression technique to variations on the original input music that match the original input music in terms of style, genre, beat and instrumentation, while varying specifics of timbre and rhythm. 

Unlike auto-regressive music models  \cite{borsos2022audiolm,agostinelli2023musiclm}, which can only perform music continuations -- using some prefix audio as a prompt, and having the model generate music that could plausibly come after it -- our approach allows the prompts to be placed anywhere. We explore a variety of prompt designs, including periodic, compression, and musically informed ones (e.g. masking on the beat). We find that our model responds well to prompts to make loops and variations, thus the name VampNet \footnote{To vamp is to repeat a short passage of music with variation.}. We make our code open source\footnote{\label{note1}\url{https://github.com/hugofloresgarcia/vampnet}} and highly encourage readers to listen to our audio samples\footnote{\label{note2}audio samples: \url{https://tinyurl.com/bdfj7rdx}}. 
 
\section{Background}
\label{sec:related_work}

Two-stage approaches to generative modeling have gained traction in image \cite{rampas2022paella, chang2022maskgit, esser2021taming, rombach2021highresolution} and audio \cite{borsos2022audiolm,agostinelli2023musiclm, borsos2023soundstorm, copet2023simple} synthesis, largely in part due to their computational efficiency. In the first stage, a discrete vocabulary of ``tokens'' is learned for the domain of interest. The input is put through an encoder to obtain these tokens, which can be converted back into the input domain via a corresponding decoder. In the second stage, a model is trained to generate tokens, and is optionally given some conditioning (e.g. previous tokens, a text description, a class label) to guide generation. 

\subsection{Stage 1: Tokenization}
In images, visual tokenization has been leveraged for state-of-the-art classification \cite{dosovitskiy2020image} and synthesis \cite{chang2022maskgit, esser2021taming, van2017neural, rombach2021highresolution}. The most popular approach is to use vector quantization on a latent space. Similar approaches have been explored for audio \cite{garbacea2019low}, but until recently such approaches have been restricted to low sampling rates (e.g. 16khz), or have been restricted to speech audio. The ``sampling rate'' of the latent space (the number of latent vectors required every second to represent audio) is a critical aspect of the tokenization scheme. The lower the sampling rate of the latent space, the easier the next stage (generation) will be to accomplish. Recently, methods based on residual vector quantization \cite{zeghidour2021soundstream, defossez2022high} have been proposed for audio tokenization at high compression rates with good reconstruction quality of high-sample-rate audio.

The primary work we leverage for audio tokenization is the Descript Audio Codec (DAC) \cite{kumar2023highfidelity}. With DAC, audio is encoded into a sequence of tokens via a fully convolutional encoder. The output of this encoder is then quantized using a hierarchical sequence of vector-quantizers \cite{van2017neural}. Each quantizer operates on the residual error of the quantizer before it. Because of this residual vector quantization, DAC is able to reconstruct audio with very high quality, at a high compression ratio. It, along with its predecessors \cite{defossez2022high, zeghidour2021soundstream}, are instrumental in enabling audio language models like AudioLM \cite{borsos2022audiolm}, MusicLM \cite{agostinelli2023musiclm}, and VALL-E \cite{wang2023neural}.  While we later briefly describe our tokenizer, the key contributions of our work are applicable to the output of any audio tokenizer and our specific audio tokenizer is not the focus of this work.

\begin{figure*}
 \includegraphics[width=\textwidth]{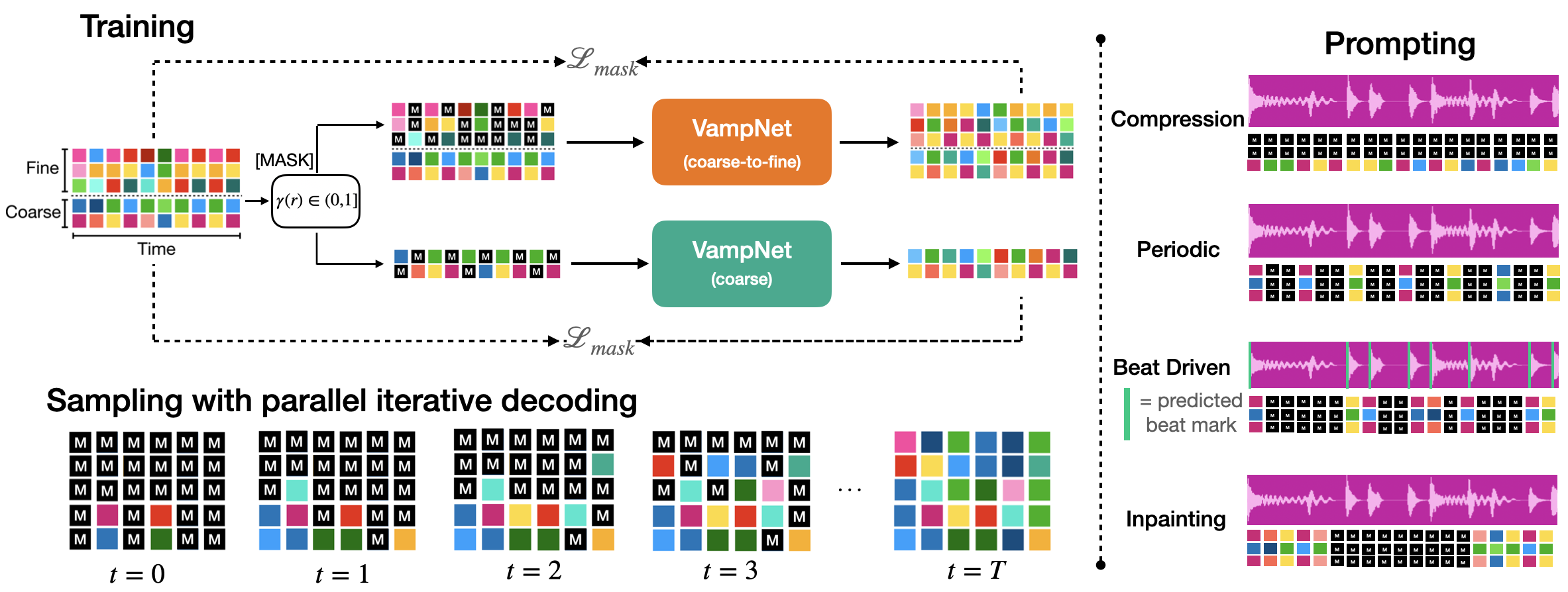}
 \caption{Training, sampling, and prompting VampNet. \textbf{Training}: we train VampNet using Masked Acoustic Token  Modeling, where we randomly mask a portion of a set of input acoustic tokens and learn to predict the masked set of tokens, using a variable masking schedule. Coarse model training masks coarse tokens. Coarse-to-fine training only masks fine tokens.  \textbf{Sampling}: we sample new sequences of acoustic tokens from VampNet using parallel iterative decoding, where we sample a subset of the most confident predicted tokens each iteration. \textbf{Prompting}: VampNet can be prompted in a number of ways to generate music. For example, it can be prompted periodically, where every $P$th timestep in an input sequence is unmasked, or in a beat-driven fashion, where the timesteps around beat markings in a song are unmasked.
}
 \label{fig:overview}
\end{figure*}

\subsection{Stage 2: Generation}
Given audio encoded as tokens, the common approach is to use an autoregressive model \cite{radford2018improving} for generation. State-of-the-art (SOTA) audio generation approaches like AudioLM \cite{borsos2022audiolm}, MusicLM \cite{agostinelli2023musiclm}, and JukeBox \cite{dhariwal2020jukebox} use this approach, generating each acoustic token in the sequence in a step-by-step fashion using transformer-based \cite{vaswani2017attention} decoder-only models. Autoregressive sampling is slow in nature due to the high number of steps required at inference time \cite{chang2022maskgit}. Further, autoregressive models inherently restrict downstream applications, as each generated token is only conditioned on the previous tokens. For an autoregressive model to perform tasks like inpainting (``filling in the middle''), one must re-arrange the data during training \cite{bavarian2022efficient}.

In language, masked modeling has been used extensively as a pre-training procedure for high-quality semantic representations \cite{devlin2018bert}. This procedure has also been extended for representation learning in images \cite{he2022masked} and audio \cite{chung2021w2v}. Masked modeling for representation learning generally has a constant mask probability. For example, in BERT \cite{devlin2018bert}, tokens are masked 15\% of the time during training. It has been shown that this approach is equivalent to a single-step discrete diffusion model \cite{austin2021structured}, that uses masking for its noising procedure. Therefore, we can extend masked modeling to masked generative modeling by varying the probability of masking a token during training. This was done for image generation in MaskGIT \cite{chang2022maskgit}, and in language \cite{austin2021structured}. Similar to diffusion modeling \cite{song2019generative, ho2020denoising}, which seeks to synthesize data starting from random noise through a series of denoising steps, masked generative modeling seeks to synthesize data starting from completely masked data through a series of ``unmasking'' steps. 

Key to the efficiency of MaskGIT and related approaches is a \textit{parallel iterative decoding procedure}. In parallel iterative decoding, the model predicts every token in the output sequence in a single forward pass. However, after just one forward pass of the model, the output often does not have high quality. The output of the first sampling step is re-masked, with a lower masking probability, and then put through the model again. In this way, masked generative models can efficiently refine their output, resulting in high quality generation.

In unconditional generation tasks, the model is asked to generate a realistic sample from the target data distribution from scratch, without any guidance. This is a difficult problem, as many target data distributions are highly multimodal. Unconditional generative models are susceptible to mode collapse \cite{srivastava2017veegan}, blurry samples, mode averaging, and other issues\cite{salimans2016improved}. Therefore, some conditioning is helpful as it provides some signal for the model to resolve the multimodality. Conditioning is also a commonly used method to guide the output of the system towards desired content.

Conditioning can take the form of a class label, a genre tag or lyrics \cite{dhariwal2020jukebox}, or an associated text description \cite{chang2023muse, rombach2021highresolution, agostinelli2023musiclm}. Conditioning can also be applied at every timestep, like the semantic tokens of AudioLM \cite{borsos2022audiolm}, or aligned text or phonemes for text-to-speech generation \cite{wang2023neural}. 


In this work,we adopt a masked generative modeling approach with a parallel iterative decoding procedure, inspired by work in vision such as \textit{MaskGIT} \cite{chang2022maskgit} and \textit{Paella} \cite{rampas2022paella}, as illustrated in Figure \ref{fig:inference}. We do not apply any conditioning beyond that provided by the unmasked tokens in our encoded audio. As we show later, different approaches to masking, applied at inference time, can be used to steer generation in useful and artistic ways.

In training, tokens are masked randomly throughout the sequence. The model is then asked to predict the value of each of the masked tokens in a single forward pass, but it is conditioned on all of the unmasked tokens, both in the future as well as in the past. We vary the number of tokens that are masked during training, allowing us to generate audio at inference time through a sampling procedure. We now describe our method in more detail. 

\section{Method}\label{sec:method}

We adapt the procedure of \textit{Masked Visual Token Modeling}, proposed in MaskGIT \cite{chang2022maskgit} to 
audio, accounting for several key differences between the vision and audio domain.
We call our approach \textit{Masked Acoustic Token Modeling}. 

\subsection{Masked Acoustic Token Modeling}

We first train an audio tokenizer based on the techniques described in DAC \cite{kumar2023highfidelity}. Unlike the visual tokens of MaskGIT, our acoustic tokens are hierarchical in nature due to residual vector quantization. 
As a first step, the audio signal $x$ is encoded at each time step $t$ as a a $D$ dimensional latent vector $Z$. We then quantize $Z$ using $N$ vector quantizers. Quantizer 1 produces $\hat{Z_1}$, a quantized approximation of $Z$ that has residual error $R_1 = Z - \hat{Z}_1$. Thereafter, the residual from each quantizer $i$ is passed to the next quantizer $i+1$, which produces a quantized approximation of the remaining residual error: $R_i \approx \hat{Z_{i+1}}$. Vector $Z$ is reconstructed by summing the output of the $N$ quantizers: $Z = \sum_{i=1}^{N} {\hat{Z_i}}$.

Since the encoded signal is represented as a quantized vector of $N$ discrete tokens at each timestep, we have $N$ tokens that can be masked or unmasked at each timestep. Rather than attempt to generate all tokens at once, we instead split the $N$ tokens into $N_c$ ``coarse'' tokens, and $N_f$ ``fine'' tokens, as in AudioLM. We then train two generative models: one that generates the fine tokens given the coarse tokens as conditioning, and one that generates the coarse tokens given a sequence of coarse tokens. To generate a sample (Figure \ref{fig:inference}), we chain the two models together. First, we apply the coarse model to generate a sequence of coarse tokens. Then, we apply the coarse-to-fine model to generate the fine tokens. We  decode the tokens to a 44.1khz waveform using the decoder of our audio tokenizer. 

\subsection{Training procedure}

Let $\mathbf{Y} \in \mathbb{R}^{T\times N}$ be a matrix representing the output of the encoder for some audio segment. Each element $y_{t,n}$ in $\mathbf{Y}$ is a token from the $n$th level codebook at timestep $t$. Let $\mathbf{Y}_M$ be the set of all masked tokens in $\mathbf{Y}$ and  $\mathbf{Y}_U$ be the set of all unmasked tokens in $\mathbf{Y}$. The model generates a probability distribution over the set of possible codebook values for each token $y \in \mathbf{Y}_{M}$, given the unmasked tokens and the model parameters $\theta$.  The training objective is to maximize the probability of the true tokens. This corresponds to minimizing the negative log likelihood.

\begin{equation}\label{eq:loss}
\mathcal{L} = - \sum_{\forall y \in \mathbf{Y}_M}{ \log p(y| \mathbf{Y}_U, \theta) \;\;\;\;   }
\end{equation}

To predict the masked tokens, we use a multi-layer bidirectional transformer, which predicts the probabilities of each possible token at every timestep, for every quantizer. If each quantizer has a codebook size of $C$ possible values, and there are $N$ quantizers, then the last layer of the network will be a fully connected layer of shape $(E, CN)$, where $E$ is the dimensionality of the output of the last layer. We then reshape this output into $(EN, C)$, and compute the cross-entropy loss between the ground-truth one-hot token and the predicted token. Because the transformer is bidirectional, it can attend to all tokens in the input sequence to optimize the loss for each token.

For the coarse-to-fine generative model, the input sequence always contains $N_c$ coarse tokens, and the masking operation is restricted to the $N_f$ fine tokens. The last layer of this network only predicts masked fine tokens. Otherwise, the training procedure for both models is identical.

\subsection{Sampling}
We follow the same iterative confidence-based sampling approach used in MaskGIT. More concretely, given $Y_M$ as the set of masked tokens and  $Y_U$ as the set of unmasked tokens,  do:

 \vspace{-2mm}
 \begin{tight_enumerate}     
    \item \textbf{Estimate.}  For each masked token $y$ in $Y_M$, estimate the conditional probability distribution over its vocabulary of codebook values $V$.
    
    \item \textbf{Sample.} For each masked token, sample from the distribution to generate an associated token estimate $\hat{y} \in V$. We don't use any sampling tricks in this step, sampling from the categorical probability distribution for each token as-is.  

    \item \textbf{Rank by Confidence.} Compute a confidence measure for each of the sampled tokens by taking their prediction log-probabilities and adding temperature-annealed Gumbel noise to them: 
    \begin{equation}\label{eq:confidence}
    confidence(\hat{y}_t) = log(p(\hat{y}_t)) + temp \cdot g_t
    \end{equation}
    where $\hat{y}_t$ is a token estimate at timestep $t$, $g_t$ is an i.i.d sample drawn from Gumbel(0,1) \cite{gumbel1954statistical}, and $temp$ is a hyperparameter that is linearly annealed to 0 over the number of sampling iterations. 
    Then, sort the set of sampled token estimates by the confidence  computed above. We find that high temperature values (e.g. $>6.0$) result in higher quality samples. 

    \item \textbf{Select.}
    Pick the number of tokens to mask at the next sampling iteration, $k$, according to the masking schedule \footnote{$k = \gamma (\frac{t}{t_T}) D$, where $t$ is the current iteration, $t_T$ is the total number of iterations, and $D$ the total number of tokens in the sequence. The scheduling function $\gamma$ is a cosine schedule.}. Take the $k$ lowest confidence estimates and toss them out, re-masking their tokens.  Place the remaining high-confidence token estimates in $Y_U$, removing their tokens from $Y_M$.
    \item \textbf{Repeat} Return  to step 1 until the number of iterations has been reached.
 \end{tight_enumerate}


\subsection{Prompting}
\label{sec:prompting}
Interactive music editing can be enabled by incorporating human guidance in the sampling procedure through the conditioning prompt of unmasked tokens. Because our approach isn't conditioned on any signal other than the input audio itself, we find that various types of prompts are useful for obtaining coherent samples, as they lower the amount of multimodality when sampling from the model. Like AudioLM, we can prompt our model with prefix audio of some duration (usually between 1 and 4 seconds), and it will provide a continuation of that audio. Unlike AudioLM, and other auto-regressive approaches, we can also prompt our model with suffix audio, and it will generate audio that leads up into that suffix. We can provide prefix and suffix audio, and the model will generate the remaining audio, such that it is appropriate, given
the specified prefix and suffix.

We can also apply a ``periodic'' prompt, where all but every $P$th timestep are masked.%
The lower $P$ is, the more the generated audio will sound like the original, as the model is highly conditioned. For example if $P = 2$, then the model is essentially behaving like a upsampler, imputing the tokens for every other timestep. As $P$ increases, the model shifts from behaving in a \textit{compression} mode to a \textit{generative} mode, creating variations that match the style of the original.

Another useful style of prompt are ``compression'' prompts, where all codebooks other than the most coarse-grained are masked.  This gives the model strong conditioning on every timestep, so the model is likely to produce audio that closely matches the original. We can combine this prompt with a periodic prompt with low $P$ for even more extreme compression ratios. Given the bitrate of the codec $B$ , which has number of codebooks $N$, a downsampling rate $P$ for the periodic prompt, and a number of kept codebooks $N_{k}$, we can achieve a bitrate of $B / P(N - N_{k})$.

Finally, we can design music-specific prompts, which exploit knowledge about the structure of the music. More concretely, we explore beat-driven prompting, where timesteps that fall on or around the beat are left unmasked. The model is left to create music between these beats, resulting in interesting variations on the original music. These prompts can all be combined to create a very useful music creation tool. In concert with a well designed user interface, VampNet shows promise as the basis for a next-generation music editing and creation suite.

\section{Experiments}
\label{sec:experiment}
Our experiments aim to evaluate VampNet's capability to both compress and generate music, given the various prompting strategies described in Section \ref{sec:prompting}. For our objective audio quality measures, we use a multiscale mel reconstruction error and the Fréchet Audio Distance (FAD). Mel-reconstruction error is defined as the $L1$ distance between log-mel spectrograms at various time-scales, 

\begin{equation}
    \text{D}_{F,M} = || \hat{S}_{F,M} - S_{F,M} ||_1
\end{equation}

where $F$ is the FFT size of each spectrogram, and $M$ is the number of mel-frequency
bins. We use $F \in [2048, 512]$ and $M \in [150, 80]$, with a hop size of $\frac{1}{4}$ the FFT size. Mel-reconstruction is valuable as a metric for compression quality, but not for generation quality, since it is likely that models produce audio that does not match one to one with the original target audio. For generation quality, we use FAD, which measures the overlap between distributions of real and generated audio. Unlike mel-reconstruction, FAD is geared more towards evaluating if sample quality falls within the data distribution of the real audio, and can be used to evaluate generation quality. 

\subsection{Dataset}
Similar to JukeBox \cite{dhariwal2020jukebox}, we collect a large dataset of popular music recordings. Our dataset consists of 797k tracks, with a sampling rate of 32 khz. These tracks are resampled to 44.1kHz to make compatible with our tokenizer. Our dataset contains music from 
thousands of artists across genres described in Echo Nest's Every Noise at Once \footnote{\url{https://everynoise.com/engenremap.html}}.

We use a subset of 2k tracks for validation, and another subset of 2k tracks for testing. We ensure that there is no artist overlap between train, validation, and test tracks. 
In addition, we collect a set of music and non-music data (speech, environmental sound), which we used to train our tokenizer, using the datasets described in DAC \cite{kumar2023highfidelity}.  
All audio is normalized to -24dbFS. We do not use any metadata about these files during training, as our model is trained unconditionally. 

\begin{figure}
 \centerline{
 \includegraphics[width=1.0\columnwidth]{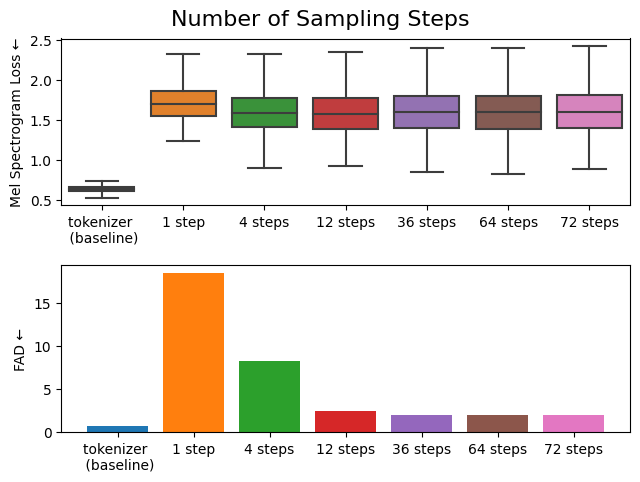}}
 \caption{Mel reconstruction error (top) and Fréchet Audio Distance (FAD, bottom) for VampNet samples taken with varying numbers of sampling steps, taken using a periodic prompt of $P=16$. The samples were generated by de-compressing tokens at an extremely low bitrate (50 bps), effectively generating  variations of the input signals.
 }
 \vspace{-5mm}
 \label{fig:sampling-steps}
\end{figure}

\subsection{Network Architecture and Hyperparameters}
 The audio tokenizer model we use takes as input 44.1kHz audio, and compresses it to a bitrate of 8kbps using 14 codebooks, with a downsampling rate of 768x. The latent space therefore is at ~57Hz, with 14 tokens to predict at every timestep. We designate 4 of these tokens as the coarse tokens, and the remaining 10 as the fine tokens. Refer to the Descript Audio Codec \cite{kumar2023highfidelity} for details on the tokenizer architecture. We train the tokenizer for 250k steps.

The VampNet architecture (for both coarse and coarse-to-fine models) consists of a bidirectional transformer \cite{vaswani2017attention} with relative attention \cite{shaw2018self} and an embedding dimension of 1280 and 20 attention heads. The coarse model has 20 attention layers, while the coarse-to-fine model has 16. 
We train the coarse and coarse-to-fine model for 1M and 500k steps, respectively. We train with the AdamW optimizer \cite{Loshchilov2017fixing} with $\beta_1$ and $\beta_2$ set to 0.9 and 0.999, respectively. We use the learning rate scheduler introduced by Vaswani et al  \cite{vaswani2017attention} with a target learning rate of 0.001 and 10k warmup steps. We use a dropout of 0.1, and a batch size of 25, with a GPU memory budget of 72GB.

\subsection{Efficiency of VampNet}
\label{sec:efficiency}

We first validate that VampNet can generate realistic music audio in a low number of steps. To do this, we run VampNet using one of our prompts (the periodic prompt, with $P = 16$) on our test set, on 10-second excerpts. We vary the number of sampling steps in $[1, 4, 8, 12, 36, 64, 72]$, and report metrics for each sampling step.

\begin{figure}
 \centerline{
 \includegraphics[width=1.0\columnwidth]{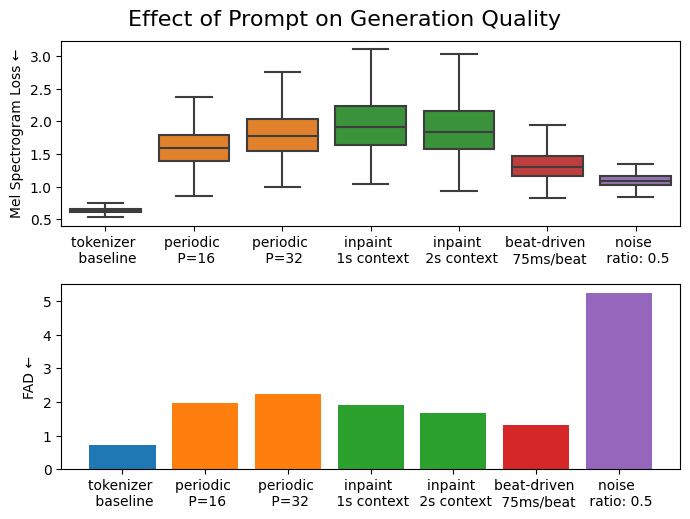}}
 \caption{ Multiscale Mel-spectrogram error (top) and Fréchet Audio Distance (FAD, bottom) for VampNet 10s samples taken with a different types of prompts. 
 }
 \label{fig:musical-sampling}
\end{figure}

\subsection{Effect of prompts}
\label{sec:exp-prompt}  

We seek to understand how VampNet responds to different prompts, as discussed in Section \ref{sec:prompting}. The prompts range from ``compression'' prompts, which compress music to a low bitrate, to more creative ``generative'' prompts. We examine whether compression and generative prompts exist on a continuum, and whether decompression from low bitrates results in generative behavior. 

We draw 2000 10-second examples from our evaluation dataset, encode them into token streams with our audio tokenizer, and manipulate the token streams in four ways:

\vspace{-2mm}
\begin{tight_enumerate}
    \item Compression prompt: $C$ codebooks are left unmasked, starting from the coarsest codebook. All other tokens are masked. We set $N_k = 1$.
    \item Periodic prompt: every $P$th timestep is left unmasked. In an unmasked timestep, tokens from every codebook are unmasked. All other tokens (e.g. tokens in timesteps that do not correspond to the period $P$) are masked. We set $P \in [8, 16, 32]$.
    \item Prefix and suffix (inpaint) prompts: a segment at the beginning and at the end of the sequence is left unmasked. All other tokens are masked. This prompt is parameterized by a context length in seconds. We set the context to be either 1 second or 2 seconds, which corresponds to 57 or 114 timesteps.
    \item Beat-driven prompt: we first process the audio waveform with a beat tracker \cite{steinmetz2021wavebeat}. Then, around each detected beat, we unmask timesteps to the right of the beat. We examine a 75ms unmasked section around each beat, which is about 4 timesteps per beat.
\end{tight_enumerate}

After manipulating the input token streams with our prompts, we generate new musical signals from these masked token streams using VampNet, and compute FAD and mel-reconstruction error between the generated signals and the  input signals from our music dataset. 
We include a noisy token stream baseline, where a portion (as dictated by mask ratio $r$) of the tokens in the input token stream are  replaced with random tokens. We also include as baseline the codec by itself, as well as the coarse-to-fine model.

Finally, we examine how these prompts can be combined - specifically the compression and periodic prompts. By manipulating the hyperparameters of these prompts ($C$ and $P$), we can shift the model behavior from compression to generation. As more timesteps are masked, the model must generate plausible musical excerpts that connect the unmasked timesteps, that may not match the input music.

\section{Results and discussion}

Results for our experiment varying the number of sampling steps used to generate samples with VampNet are shown on Figure \ref{fig:sampling-steps}. We find that VampNet achieves the lowest FAD with 36 sampling steps, although 12 sampling steps achieves comparable performance. In practice, we find that samples taken with 24 steps achieve a fair trade-off between generation quality and compute speed, with 10-second samples taking around 6 seconds to sample on an NVIDIA RTX3090. In contrast, to generate 10 seconds of audio with an autoregressive model would require 574 steps, which would take around 1 min to generate 10 seconds of audio, given an autoregressive model with the same number of parameters as ours, and the same  tokenizer.

Results for our study on the effect of each prompt are shown in Figure \ref{fig:musical-sampling}. First, we note that while the noisy token baseline has comparable mel reconstruction to all prompts, it performs very poorly in terms of FAD. This indicates that while our prompting strategies may result in audio that is not a perfect match to the original input audio, it still falls inside the distribution of plausible music.

Of our proposed prompts, we find that beat-driven prompts perform best, achieving the lowest FAD of all prompts.  A notable result here is that the periodic prompt with $P=16$ (35 conditioning timesteps) performs on par with inpainting with 1 second of context (57 conditioning timesteps). Therefore, prompt techniques that spread out the conditioning tokens throughout the sequence (periodic prompts) are able to use fewer conditioning timesteps to generate samples of comparable quality to those generated by sampling techniques that place all of the conditioning tokens at the start and end of the sequences (inpainting).

\begin{figure}
 \centerline{
 \includegraphics[width=1.0\columnwidth]{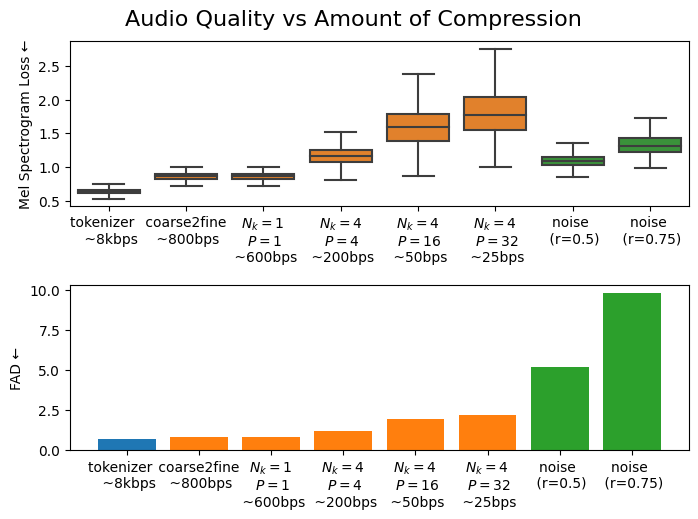}} \caption{Mel-spectrogram error (top) and Fréchet Audio Distance (FAD) (bottom) for VampNet samples at varying bitrates. A baseline is provided by replacing tokens in the input sequence with random tokens, per noise ratio $r$.}
 \label{fig:gen-compression}
\vspace{-5mm}
\end{figure}

Qualitatively, we also find that beat-driven prompts can keep a steadier tempo than other prompts, though their outputs tend to resemble the original music closer than periodic prompts. In practice, a mix of beat-driven, periodic, and inpainting prompts can be employed to steer of VampNet in creative ways. To illustrate, we highly encourage the reader to listen to the accompanying sound samples \footnote{audio samples: \url{https://tinyurl.com/bdfj7rdx}}.

We then combined periodic and compression prompting to show how the model's behavior shifts between reconstruction and generation tasks, as more tokens are masked away. 
Results for this experiment are shown in Figure \ref{fig:gen-compression}. At higher bitrates, (600 bps and above), VampNet is able to accurately reconstruct the original music signal, achieving low mel-spectrogram error and FAD values with respect to the evaluation music audio. At bitrates of 200bps and below, VampNet has comparable reconstruction quality to the noisy token baselines, indicating that the sampled VampNet signals no longer resemble the input audio in terms of fine-grained spectral structure. However, the FAD for VampNet samples at low bitrates is much lower than the FAD for noisy baselines. This indicates that even though VampNet isn't able to reconstruct the input music signal at low bitrates, it is still able to generate coherent audio signals with musical structure, that are closer to the distribution of ``real music'' than our noisy baseline.

\vspace{-5mm}
\section{Conclusion}\label{sec:conclusion}


We introduced VampNet, a masked acoustic token modeling approach to music generation. VampNet is bidirectional, and can be prompted a variety of ways using an input audio file. Through different prompting techniques, VampNet can operate in a continuum between music compression and generation, and is an excellent tool for generating variations on a piece of music. 
With VampNet, a musician could record a short loop, feed it into VampNet, and have VampNet create musical variations on the recorded idea every time the looped region repeats. 
In future work, we hope to investigate the interactive music co-creation potential of VampNet and its prompting techniques, as well as explore the representation learning capabilities of masked acoustic token modeling. 


\bibliography{ISMIRtemplate}

%
%
%
%
%

\end{document}